\documentclass[aps,pra,twocolumn,showpacs,superscriptaddress,10pt]{revtex4-1}

\usepackage[T1]{fontenc} %
\usepackage{times} %
\usepackage{color,graphicx} %
\usepackage{subfig,array} %
\usepackage{amsthm,amssymb,amsmath} %
\usepackage[breaklinks]{hyperref} %
\usepackage[singlelinecheck=off,justification=raggedright]{caption} %

\hypersetup{colorlinks=true, linkcolor=blue, citecolor=red, urlcolor=cyan}

\newcommand\ket[1]{\ensuremath{|#1\rangle}} %
\newcommand\tr{\mathop{\rm tr}\nolimits} %

\newcommand{\nc}{\newcommand} %
\nc{\cA}{{\cal A}} \nc{\cB}{{\cal B}} \nc{\cC}{{\cal C}} %
\nc{\cD}{{\cal D}} \nc{\cE}{{\cal E}} \nc{\cF}{{\cal F}} %
\nc{\cG}{{\cal G}} \nc{\cH}{{\cal H}} \nc{\cI}{{\cal I}} %
\nc{\cJ}{{\cal J}} \nc{\cK}{{\cal K}} \nc{\cL}{{\cal L}} %
\nc{\cM}{{\cal M}} \nc{\cN}{{\cal N}} \nc{\cO}{{\cal O}} %
\nc{\cP}{{\cal P}} \nc{\cQ}{{\cal Q}} \nc{\cR}{{\cal R}} %
\nc{\cS}{{\cal S}} \nc{\cT}{{\cal T}} \nc{\cU}{{\cal U}} %
\nc{\cV}{{\cal V}} \nc{\cW}{{\cal W}} \nc{\cX}{{\cal X}} %
\nc{\cZ}{{\cal Z}}

\begin{document}


\title{Physical origins of ruled surfaces on the reduced density
  matrices geometry}

\author{Ji-Yao Chen}%
\affiliation{State Key Laboratory of Low Dimensional Quantum Physics,
  Department of Physics, Tsinghua University, Beijing, China}%
\affiliation{Perimeter Institute for Theoretical Physics, Waterloo,
  Ontario, Canada}%
\author{Zhengfeng Ji}%
\affiliation{Centre for Quantum Computation \& Intelligent Systems,
  School of Software, Faculty of Engineering and Information
  Technology, University of Technology Sydney, Sydney, Australia}%
\affiliation{State Key Laboratory of Computer Science, Institute of
  Software, Chinese Academy of Sciences, Beijing, China}%
\author{Zheng-Xin Liu}%
\affiliation{Department of Physics, Renmin University of China,
  Beijing, China}%
\author{Xiaofei Qi}%
\affiliation{Department of Mathematics, Shanxi University, Taiyuan,
  Shanxi, China}%
\author{Nengkun Yu}%
\affiliation{Institute for Quantum Computing, University of Waterloo,
  Waterloo, Ontario, Canada}%
\affiliation{Centre for Quantum Computation \& Intelligent Systems,
  School of Software, Faculty of Engineering and Information
  Technology, University of Technology Sydney, Sydney, Australia}%
\affiliation{Department of
  Mathematics \& Statistics, University of Guelph, Guelph, Ontario,
  Canada}%
\author{Bei Zeng}%
\affiliation{Department of Mathematics \& Statistics, University of
  Guelph, Guelph, Ontario, Canada}%
\affiliation{Institute for Quantum Computing, University of Waterloo,
  Waterloo, Ontario, Canada}%
\author{Duanlu Zhou}%
\affiliation{Institute of Physics, Chinese Academy of Sciences,
  Beijing, China}%

\begin{abstract}
  The reduced density matrices (RDMs) of many-body quantum states form
  a convex set. The boundary of low dimensional projections of this
  convex set may exhibit nontrivial geometry such as ruled surfaces.
  In this paper, we study the physical origins of these ruled surfaces
  for bosonic systems. The emergence of ruled surfaces was recently
  proposed as signatures of symmetry-breaking phase. We
  show that, apart from being signatures of symmetry-breaking, ruled
  surfaces can also be the consequence of gapless quantum systems by
  demonstrating an explicit example in terms of a two-mode Ising
  model. Our analysis was largely simplified by the quantum de
  Finetti's theorem---in the limit of large system size, these RDMs
  are the convex set of all the symmetric separable states. To
  distinguish ruled surfaces originated from gapless systems from
  those caused by symmetry-breaking, we propose to use the finite size
  scaling method for the corresponding geometry. This method is then applied
  to the two-mode XY model, successfully identifying a ruled surface
  as the consequence of gapless systems.
\end{abstract}

\date{\today}


\maketitle

\section{Introduction}

Natural interactions in a many-body system usually involve only a few
particles. For an $N$-particle system, the Hamiltonian of the system
adopts the form $H=\sum_j H_j$ with each $H_j$ acting nontrivially on,
in most cases, only two particles. For any quantum state
$\ket{\psi^N}$ of the system, its energy can then be determined by its
two-particle reduced density matrices ($2$-RDMs). Consequently, the
ground state energy of the system can be solely read out from
the $2$-RDMs of the ground state $\ket{\psi_0^N}$.

For a Hamiltonian $H(\vec{\lambda})$ containing some set of parameters
$\vec{\lambda}$, the ground state $\ket{\psi_0^N(\vec{\lambda})}$ may
change suddenly while the parameter $\vec{\lambda}$ changes smoothly,
leading to a quantum phase transition. Such a change can also be
captured by the $2$-RDMs of $\ket{\psi_0^N(\vec{\lambda})}$, which is
reflected by a sudden change in the set of all the $2$-RDMs, which
is known to be convex. It is then highly desired to characterize such
changes geometrically on the set of $2$-RDMs.

However, the geometric shape of the set of $2$-RDMs, denoted by
$\Theta_2^N$, is notoriously hard to characterize in general, apart
from the obvious fact that $\Theta_2^N$ is a convex set. How to
characterize $\Theta_2^N$ has been a central topic of research in the
quantum marginal problem and the $N$-representability problem, which
dates back to the
1960s~\cite{Col63,Erd72,klyachko2006quantum,EJ00,SM09}. Recently, it
has been shown that the characterization of $\Theta_2^N$ is a hard
problem even with the existence of a quantum
computer~\cite{Liu06,LCV07,WMN10}. Nevertheless, many practical
approaches are developed to characterize the properties of the set,
and to retrieve useful information that characterizes the physical
properties of the system~\cite{VC06,GM06}.

Among these approaches, one important idea is to study the $2$ and $3$
dimensional projections of these $2$-RDMs~\cite{EJ00,VC06,GM06,SM09},
such that the properties of the different quantum phases are visually
available. It has been shown that a flat portion of the
$2$-dimensional projection can already signal first-order phase
transitions~\cite{chen2015discontinuity,zauner2014symmetry}. However,
for continuous phase transitions, $2$-dimensional projections contain
no information, and one needs to further examine $3$-dimensional
projections.

It is observed that the emergence of ruled surfaces on the boundary of
the $3$-dimensional projections of the $2$-RDMs signatures
symmetry-breaking phase~\cite{zauner2014symmetry}. With a
generalization, the ruled surfaces can also signal the symmetry
protected topological phase~\cite{chen2016geometry}. And it is
interesting to note that the connection of ruled
surfaces on the boundary of certain convex body and phase transitions
dates back to Gibbs in the 1870's~\cite{Gib73a,Gib73b,Gib75,Isr79}.
The convex bodies under consideration in Gibbs' original work are in
the context of classical thermodynamics, and the case of quantum
many-body physics is rather different. It nevertheless indicates that
the convex geometry approach is a fundamental and universal idea.

It remains unclear whether there are other physical mechanisms that
may lead to the emergence of ruled surfaces on the boundary of the
convex set of RDMs. We give an affirmative answer in this work and
show that ruled surfaces can also be a consequence of gapless
systems. The underlying idea is simple: if the Hamiltonian
$H(\vec{\lambda})$ is gapless for some continuous region of the
parameter $\vec{\lambda}$, then for each $\vec{\lambda}$, the
corresponding low energy states may be projected onto a line on the
boundary of the $3$-dimensional surface, and the continuous changes of
$\vec{\lambda}$ then result in a ruled surface.

We demonstrate our ideas by studying various models of many-body
bosonic systems. The choice of such systems are due to their
simplicity to analyze. First of all, due to the exchange symmetry of
bosons, the wave function $\ket{\psi^N}$ of the system is
symmetric and consequently all the $2$-RDMs are in fact the same. We
then denote such a $2$-RDM by $\rho^N_2$.
Furthermore, due to the quantum de Finetti's theorem, in the large $N$
limit, $\rho^N_2$ has a relatively simple description, which is
exactly the set of all separable $2$-particle density matrices. For
low dimensional single particle Hilbert space, e.g. two-mode bosonic system,
this then leads to analytical results to the
$3$-dimensional projections of the set of all $\rho^N_2$. This allows
us to analyze the ruled surfaces on the boundary and their
originality.

We study the two-mode Ising model in detail to show a ruled surface
that is a direct consequence of gapless systems. We argue that
such kind of ruled surfaces are in fact quite a common
phenomenon in bosonic systems, since bosons are `inclined' to be
gapless -- for any entangled ground states, two-particle correlation
functions cannot decay exponentially with any distance defined, due to
the symmetry of the system. To distinguish ruled surfaces originated
from gapless systems from those from symmetry-breaking solely through the
geometry of RDMs, we propose to use finite system size scaling of the
corresponding geometry. We apply this finite size scaling method to the
two-mode XY model, to identify a ruled surface as a consequence of
gapless systems.

\section{Background and Notations}

In this section we recall the quantum finite de Finetti's theorem and
its consequences on the RDMs. We start with considering an $r$-mode
bosonic system of $N$ bosons with the single-particle Hilbert space
$\mathcal{H}$. For our purpose we assume the dimension of
$\mathcal{H}$ is finite. Define the collective spin operators of $N$
spins to be
\begin{equation}
  J^N_x=\sum_{i=1}^N S_x^i,\ J^N_y=\sum_{i=1}^N S_y^i,\ J^N_z=\sum_{i=1}^N S_z^i.
\end{equation}
Here $\vec{S}$ is spin operator for spin-${(r-1)/2}$.

We consider Hamiltonians with two-body interaction in terms of
$J^N_x,J^N_y,J^N_z$. More precisely, there is in fact a sequence
of Hamiltonians for different system size $N$, each of which is denoted by
$H^N$. We focus on systems that approach the large system size limit (i.e.
$N\rightarrow\infty$). The celebrated quantum de Finetti's theorem
states the
following~\cite{stormer1969symmetric,hudson1976locally,lewin2014derivation}:
\textit{ For any $N$-boson wave function $\ket{\Psi_N}$ that lies in
  the symmetric subspace of $\mathcal{H}^{\otimes N}$, and for any
  constant integer $k>0$ that is independent of $N$, the $k$-RDM
  $\rho_k$ of $\ket{\Psi_N}$ is a mixture of product states of the
  form $\ket{\alpha}^{\otimes k}$, in the $N\rightarrow\infty$ limit.}

For a two-body Hamiltonian $H^N$, the ground state energy is
determined by the $2$-RDM $\rho^N_2$ (of the $N$-particle wave
function $\ket{\Psi_N}$), i.e.
\begin{equation}
  \label{eq:groundstate}
  E^N_0={\min_{\rho^N_2}}\tr(H^N\rho^N_2).
\end{equation}

According to the quantum de Finetti's theorem, $\rho^{\infty}_2$ is
separable. The set of all $\rho^{\infty}_2$ is convex, denoted by
$\Theta^{\infty}_2$ with the extreme points
$\ket{\alpha}\ket{\alpha}$. Therefore in the large $N$ limit,
Eq.~\eqref{eq:groundstate} equals exactly the Hartree's mean field
energy. This fact is independent of the details of the Hamiltonian.

We now consider the Hamiltonians with parameters
$\vec{\lambda}=(\lambda_0,\lambda_1,\lambda_2)$, i.e.
\begin{equation}
  \label{eq:HN}
  H^N(\vec{\lambda})=\sum_{i=0}^2\lambda_i f_i(N)H^N_i,
\end{equation}
where each $H^N_i$ denotes single particle or two-body interaction in
terms of $J^N_x,J^N_y,J^N_z$ for a system of size
$N$, 
and $f_i(N)$ is a scaling factor to make energy per particle bounded
and meaningful in the large $N$ limit. Explicitly, we choose
$f_i(N)=1$ for single particle terms, and $f_i(N)=\frac{1}{N}$ for
two-body interaction terms. The set
\begin{eqnarray}
  \label{eq:ThetaN}
  \Theta^N_2(H^N)=\{(x,y,z) |\rho^N_2\in\Theta^N_2\}
\end{eqnarray}
is a three-dimensional projection of $\Theta^N_2$, where
\begin{eqnarray}
  \label{eq:xyz}
  x&=&\frac{f_0(N)}{N}\tr\left(\rho^N_2H_0^N\right),\nonumber\\
  y&=&\frac{f_1(N)}{N}\tr\left(\rho^N_2H_1^N\right),\\
  z&=&\frac{f_2(N)}{N}\tr\left(\rho^N_2H_2^N\right).\nonumber
\end{eqnarray}
And $H^N(\vec{\lambda})$ corresponds to the supporting hyperplane of
$\Theta^N_2(H^N)$, i.e., a parameter vector $\vec{\lambda}$ gives a
normal vector of one supporting hyperplane of $\Theta^N_2(H^N)$. For
any $\vec{\alpha}\in\Theta^N_2(H^N)$,
$\vec{\alpha}\cdot\vec{\lambda}\geq E^N_0(\vec{\lambda})/N$. We also
denote that when $N\to\infty$,
\begin{eqnarray}
  \Theta^{\infty}_2(H^{\infty})=\{(\bar{x},\bar{y},\bar{z}) |\rho^{\infty}_2\in\Theta^{\infty}_2\},
\end{eqnarray}
where $(\bar{x},\bar{y},\bar{z})$ is the corresponding limit of
$(x,y,z)$.

We are interested in the geometry of $\Theta^N_2(H^N)$ and its
relation with physical properties of the system $H^N(\vec{\lambda})$, especially those related with quantum phase and
phase transition.

\section{The two-mode Ising model}
\label{sec:Ising}

We start to examine the geometry of $\Theta^N_2(H^N)$ for the two-mode
Ising model, where we take the spin operators as the spin-$1/2$ Pauli operators
for convenience, e.g. $S_x=X,S_y=Y,S_z=Z$. The Hamiltonian reads
\begin{equation}
  H^N_{\mathrm{Ising}}=\frac{J}{N} (J^N_x)^2+B_z J^N_z+B_x J^N_x,
\end{equation}
where an extra $B_xJ^N_x$ term has been added to explicitly break the $Z_2$ symmetry in the traditional transverse Ising model when $B_x\neq 0$. This term is chosen for the reason that it corresponds to the order parameter of the $Z_2$ symmetry-breaking phase of the transverse Ising model~\cite{zauner2014symmetry}.

The corresponding

\begin{equation}
\label{Theta_Ising}
  \Theta^N_2(H^N_{\mathrm{Ising}})=\{(x, y, z) |\rho^N_2\in\Theta^N_2\}
\end{equation}

is given by

\begin{eqnarray}
  \label{eq:Ising}
  x &=& \frac{1}{N^2}\tr\left(\rho^N_2 (J^N_x)^2\right) \nonumber\\
    &=& \tr\left(\rho^N_2(\frac{1}{N}I+\frac{N-1}{N}X\otimes X)\right),\nonumber\\
  y &=& \frac{1}{N}\tr\left(\rho^N_2 J^N_z\right) = \tr\left(\rho^N_2Z\otimes I\right),\\
  z &=& \frac{1}{N}\tr\left(\rho^N_2 J^N_x\right) = \tr\left(\rho^N_2X\otimes I\right).\nonumber
\end{eqnarray}

Here by for any two-body operator $M$, by $\tr(\rho_2^N M)$
we mean $\langle \Psi_N| M|\Psi_N\rangle$. In other words,
since all the $2$-RDMs of $\ket{\Psi_N}$ are the same,
we simply denote it by $\rho_2^N$ and do not specify which
two particles $\rho_2^N$ is acting on. Without confusion we will use this convention
throughout the paper.

\subsection{Large $N$ limit}

In the large $N$ limit, Eq.~\eqref{Theta_Ising},~\eqref{eq:Ising} is equivalent to


\begin{eqnarray}
  \Theta^{\infty}_2(H^{\infty}_{\mathrm{Ising}})=\{(\bar{x},\bar{y},\bar{z})|\rho^{\infty}_2\in\Theta^{\infty}_2\},
\end{eqnarray}
with
\begin{eqnarray}
  \bar{x}&=&\tr(\rho^{\infty}_2(X\otimes X)),\nonumber\\
  \ \bar{y}&=&\tr(\rho^{\infty}_2(Z\otimes I)),\\
  \ \bar{z}&=&\tr(\rho^{\infty}_2(X\otimes I)).\nonumber
\end{eqnarray}


The extreme points of $ \Theta^{\infty}_2(H^{\infty}_{\mathrm{Ising}})$
are given by
\begin{equation}\label{relation_1}
  \bar{x} = \bar{z}^2, ~\bar{y}^2 + \bar{z}^2 = 1.
\end{equation}

The boundary surface of
$\Theta^{\infty}_2(H^{\infty}_{\mathrm{Ising}})$ is then given by

\begin{equation}\label{boundary_1}
  \bar{x} = \bar{z}^2, \mathrm{for}~~ \bar{y}^2+\bar{z}^2\leq 1,
\end{equation}
and
\begin{equation}\label{boundary_2}
  \bar{x}+\bar{y}^2=1, \mathrm{for}~~\bar{y}^2+\bar{z}^2\leq 1.
\end{equation}

And the corresponding supporting hyperplanes are
\begin{equation}\label{hyperplane_1}
  \bar{x} + \bar{x}_0 - 2\bar{z}_0\bar{z}=0,
\end{equation}
and
\begin{equation}\label{hyperplane_2}
  \bar{x}+\bar{x}_0+2\bar{y}_0\bar{y}=2.
\end{equation}

We observe that there are two ruled surfaces
on the boundary of $\Theta^{\infty}_2(H^{\infty}_{\mathrm{Ising}})$.
For any point $(x_0,y_0,z_0)$ living on
\begin{equation}\nonumber
  \bar{x} = \bar{z}^2, \mathrm{for}~~ \bar{y}^2+\bar{z}^2\leq 1,
\end{equation}
we have part of the line $(x_0,y,z_0)$ living on the surface
of $\Theta^{\infty}_2(H^{\infty}_{\mathrm{Ising}})$. These points
give one ruled surface.
For point $(x_0,y_0,z_0)$ lives on
\begin{equation}\nonumber
\bar{x}+\bar{y}^2=1, \mathrm{for}~~\bar{y}^2+\bar{z}^2\leq 1,
\end{equation}
we have part of the line $(x_0,y_0,z)$ living on  the surface
of $\Theta^{\infty}_2(H^{\infty}_{\mathrm{Ising}})$.
These points give the other ruled surface.

\begin{figure}[!h]
  \centering
  \includegraphics[width=80mm,height=65mm]{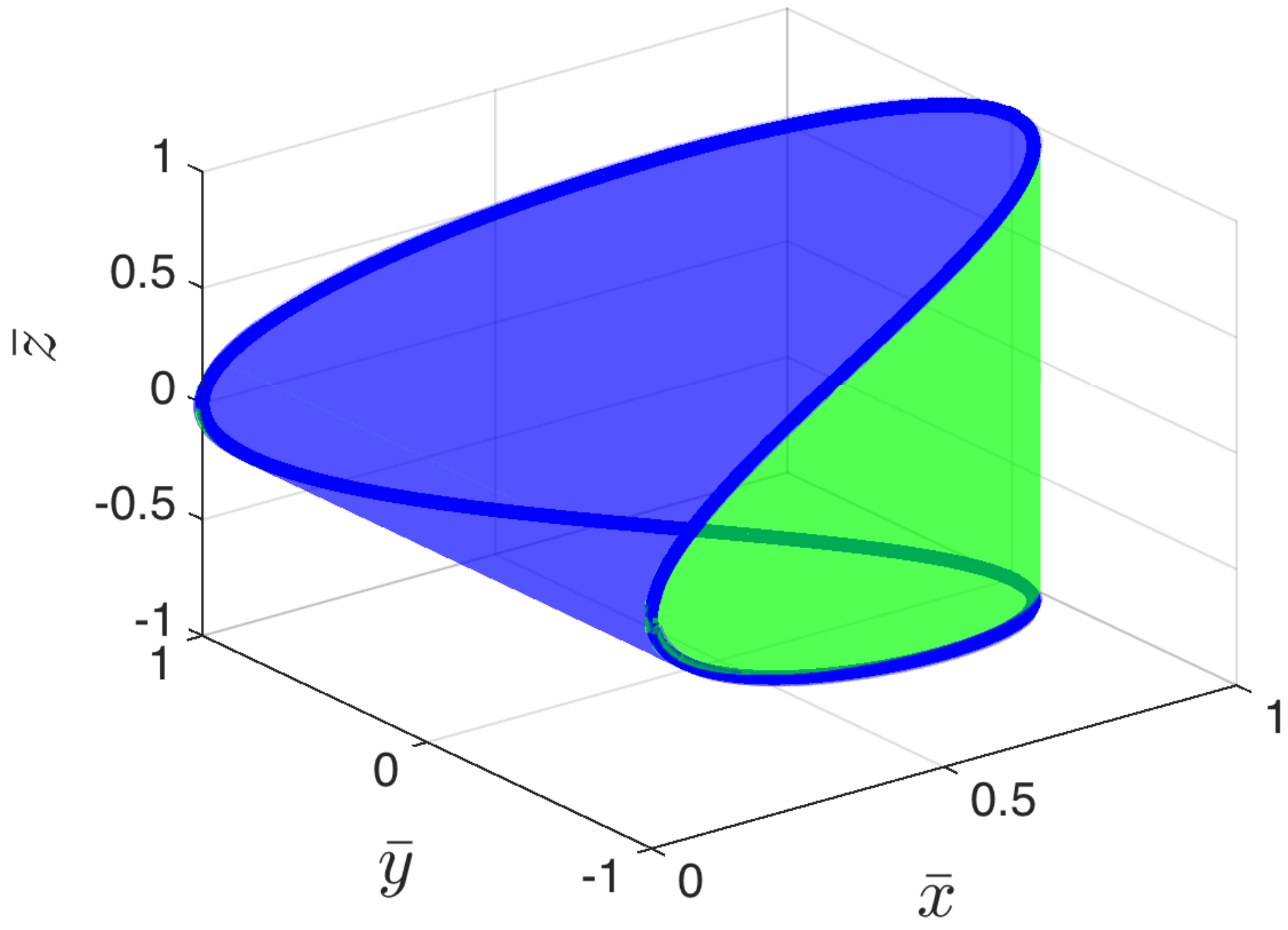}
  \caption{Convex set for Ising model in the large $N$ limit. It is
    determined by Eq.~\eqref{boundary_1},~\eqref{boundary_2}, with
    $\bar{x}=\tr(\rho^{\infty}_2(X\otimes X))$, $\bar{y}=\tr(\rho^{\infty}_2(Z\otimes I))$, $\bar{z}=\tr(\rho^{\infty}_2(X\otimes I))$.
   The green ruled surface is due to symmetry breaking,
    while the blue ruled surface is due to the fact that the system is gapless in that region.}
  \label{Ising_infty}
\end{figure}

We show the convex set
$\Theta^{\infty}_2(H^{\infty}_{\mathrm{Ising}})$ in Fig.~\ref{Ising_infty}.
There are two ruled surfaces: the blue one and the green one.
Geometrically, these two surfaces have exactly the same shape.
However, their physical origins are very different. The Hamiltonian
$H_{\mathrm{Ising}}$ is known to have a symmetry-breaking phase for $J=-1,B_x=0,|B_z|<2$,
and with $J=1, |B_x|<2$ the system is gapless~\cite{leggett2001bose}.
Therefore, the green ruled surface is due to symmetry
breaking when $J=-1,B_x=0,|B_z|<2$, while the blue ruled surface is due to that the
system is gapless when $J=1, |B_x|<2$.

When $J<0$, for simplicity, we fix $J=-1$. When $B_x = 0$, there is a
$\mathbb{Z}_2$ symmetry generated by
$O = Z_1\otimes Z_2\otimes ... \otimes Z_N$. In the range
$B_z\in [-2,2]$, the ground state is two fold degenerate (can be seen
from the energy), and a corresponding ruled surface emerge.
The phase transition here is Ising type, which can then be explained by mean
filed theory (i.e., one only needs to consider separable states, as
given in the the calculation of $\Theta^{\infty}_2(H^{\infty}_{\mathrm{Ising}})$).

\subsection{Finite size scaling}

\begin{figure*}[hbpt]
  \centering \subfloat[]{
    \includegraphics[width=80mm,height=65mm]{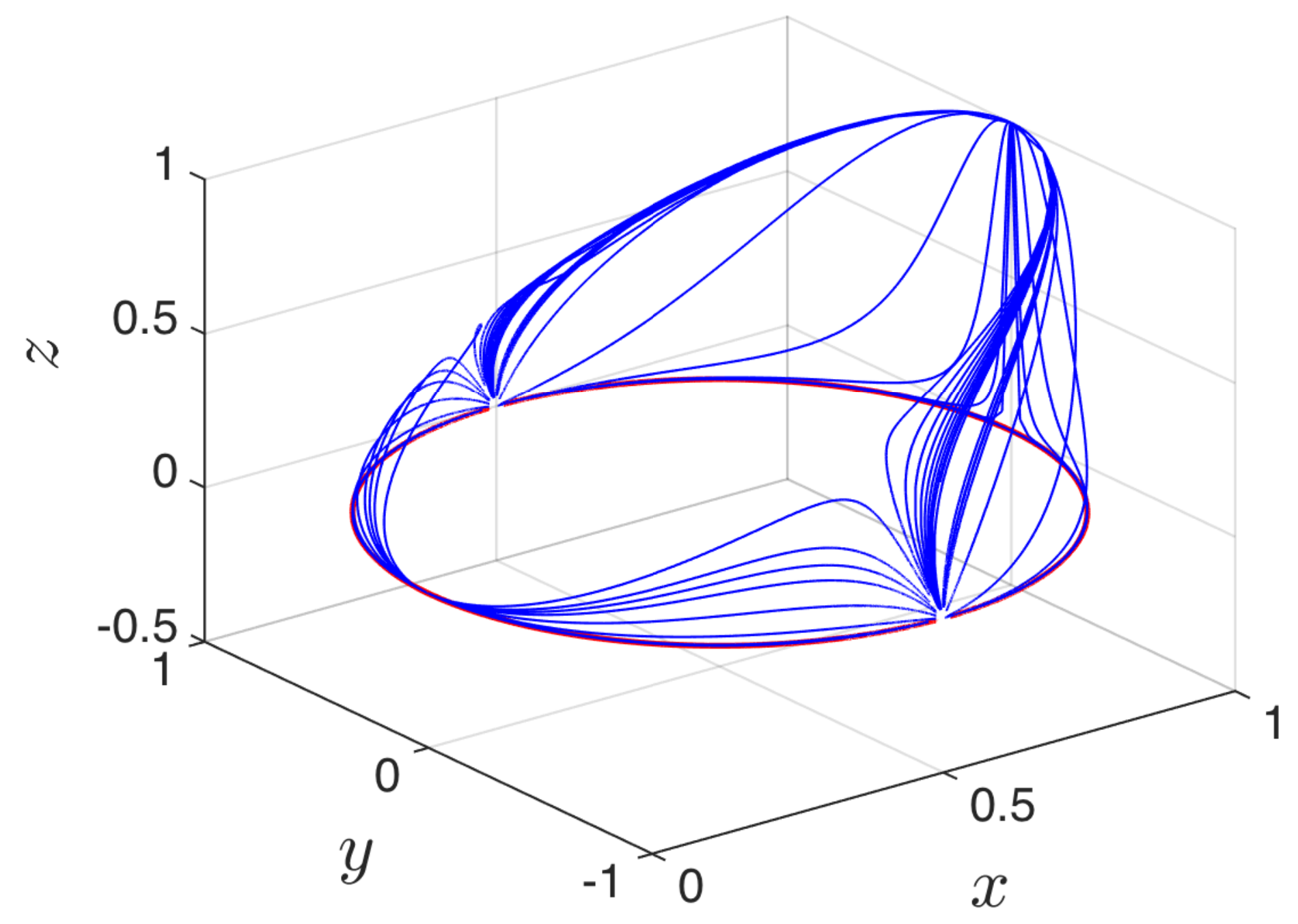}}
  \subfloat[]{
    \includegraphics[width=80mm,height=65mm]{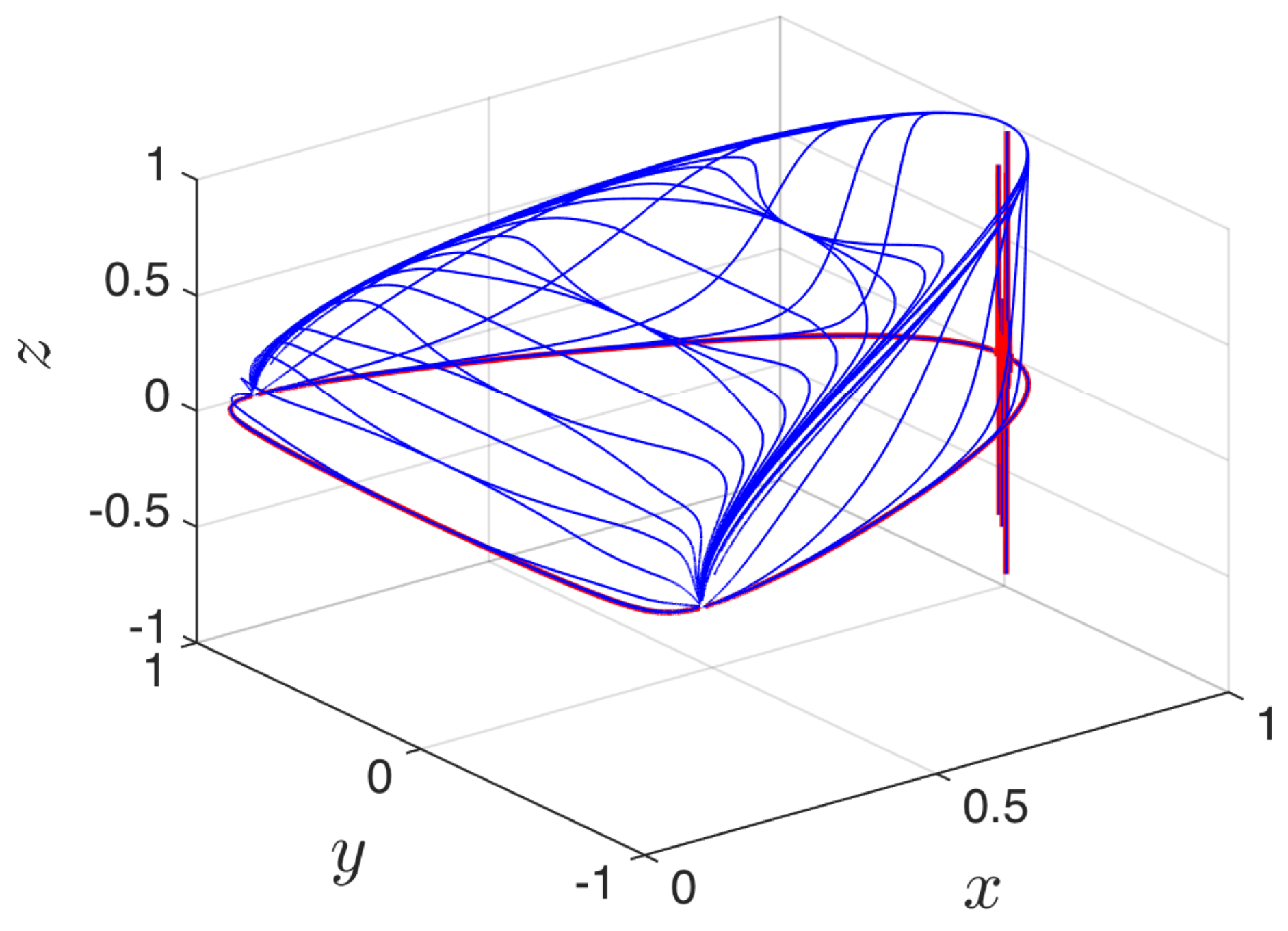}}\\
  \subfloat[]{
    \includegraphics[width=80mm,height=65mm]{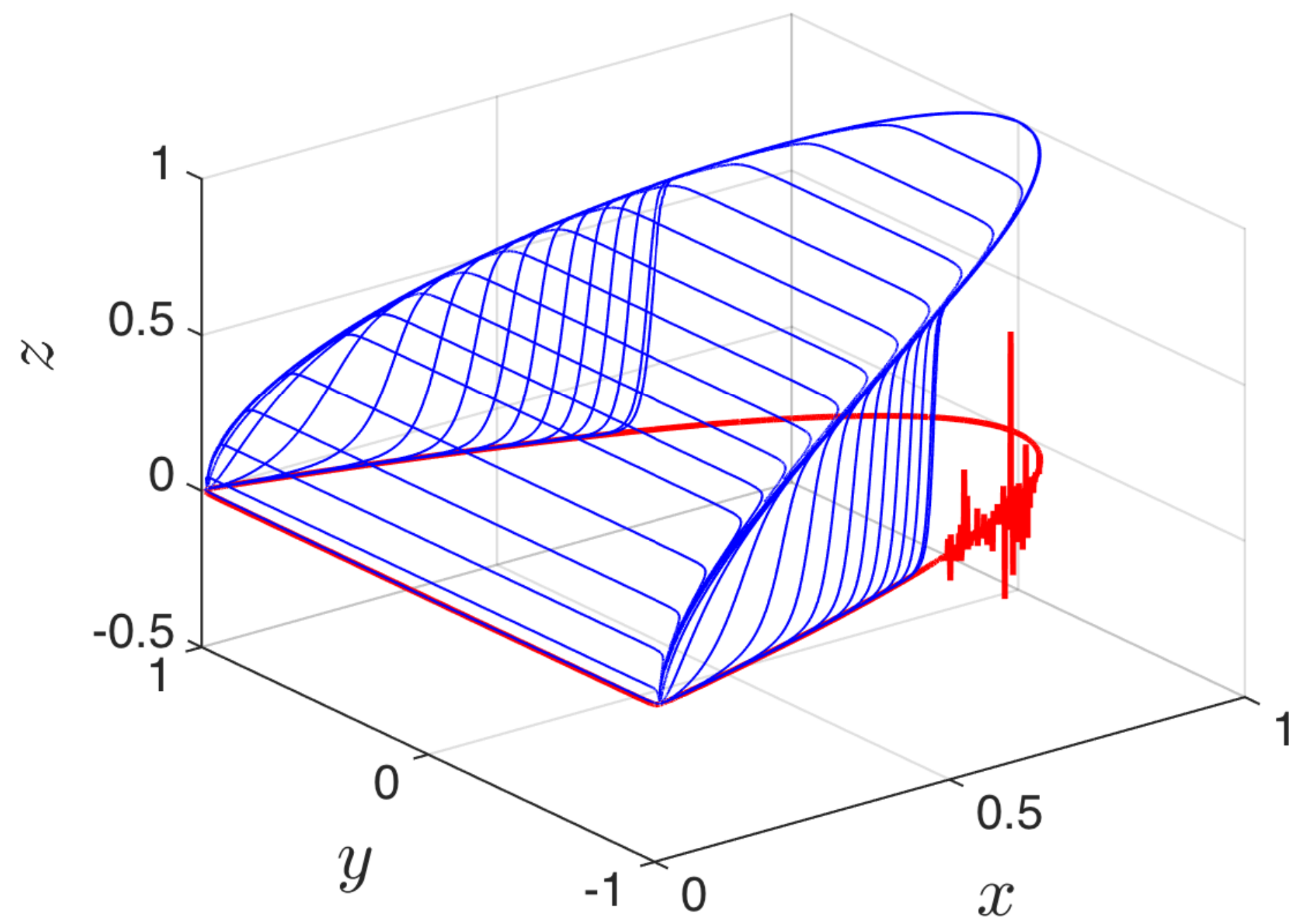}}
  \subfloat[]{
    \includegraphics[width=80mm,height=65mm]{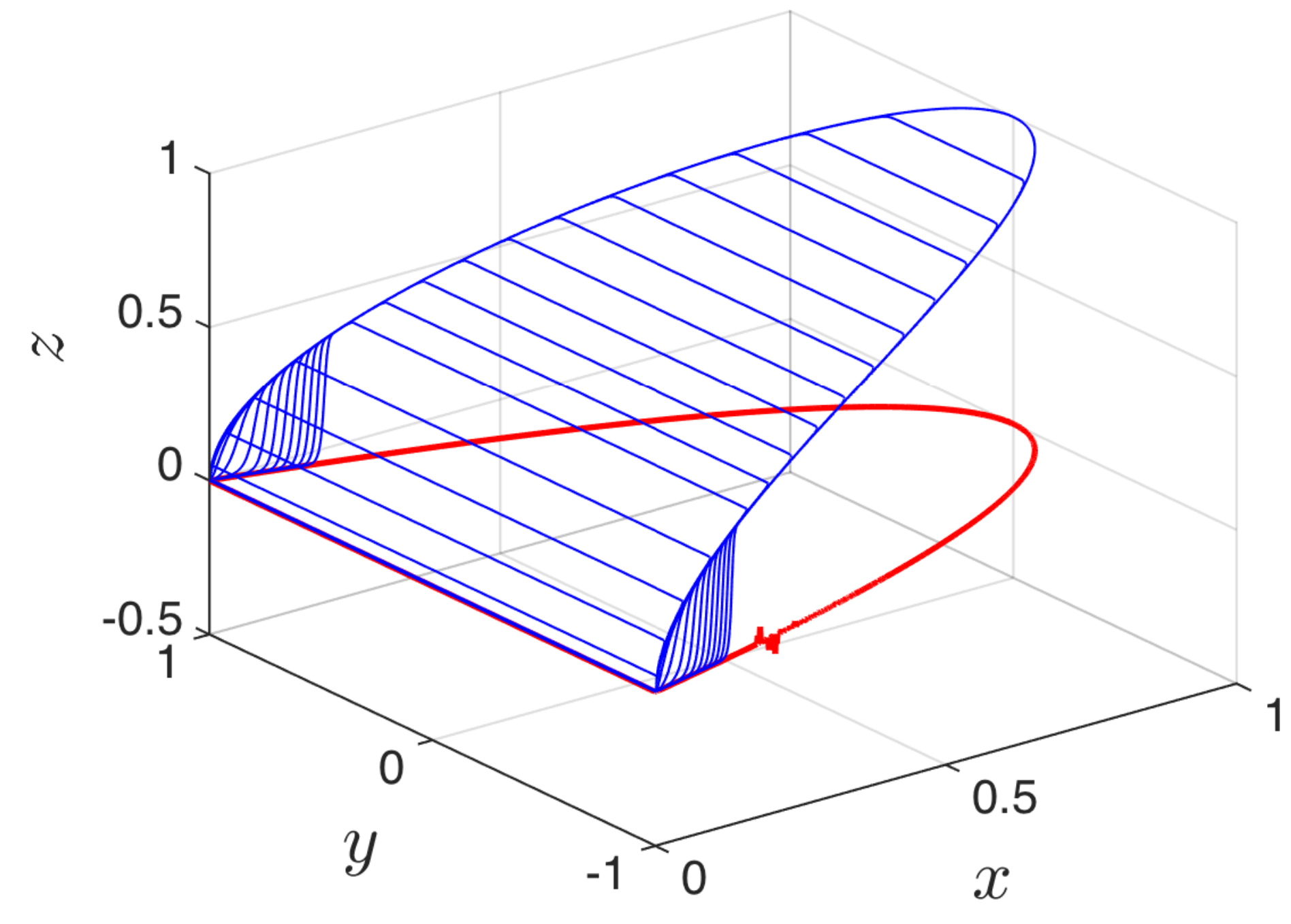}}\\
  \caption{Convex set $\Theta^{N}_2(H^{N}_{\mathrm{Ising}})$ for two-mode Ising case, with
    $x=\tr\left(\rho^N_2(\frac{1}{N}I+\frac{N-1}{N}X\otimes X)\right)$,
  $y=\tr\left(\rho^N_2Z\otimes I\right)$,
  $z=\tr\left(\rho^N_2X\otimes I\right)$.
   For clarity, we only show $B_x\le 0$ part. The result is obtained by exact
    diagonalization but with different particle numbers. The particle
    number is $N=2,10,10^2, 10^3$ for (a), (b), (c), (d) respectively.
    The ruled surface becomes more and more clear with increasing
    system size. }
  \label{Ising}
\end{figure*}

Although the two ruled surfaces have exactly the same shape for
$N\rightarrow\infty$, their different physical origins can be seen from
the finite size scaling. The finite $N$ scaling is shown in
Fig.~\ref{Ising}. Clearly, when $N$ becomes larger, the convex set
will go to the $N\rightarrow\infty$ limit which is shown in Fig.~\ref{Ising_infty}.
Due to symmetry we only show the upper part.

Equation~\eqref{relation_1} can also be verified by the large $N$ data.
We remark that, in Fig.~\ref{Ising}, there are some special points in the
$\Theta^{N}_2(H^{N}_{\mathrm{Ising}})$ for finite $N$. For example the point with parameter
$J=1, B_{x}=-1, B_{z}=0$. This point seems to be discontinuous from its
neighbor. 
These special points will become normal in the $N\to \infty$ limit.

\begin{figure}[hbpt]
  \centering
  \subfloat[]{
    \includegraphics[width=70mm,height=60mm]{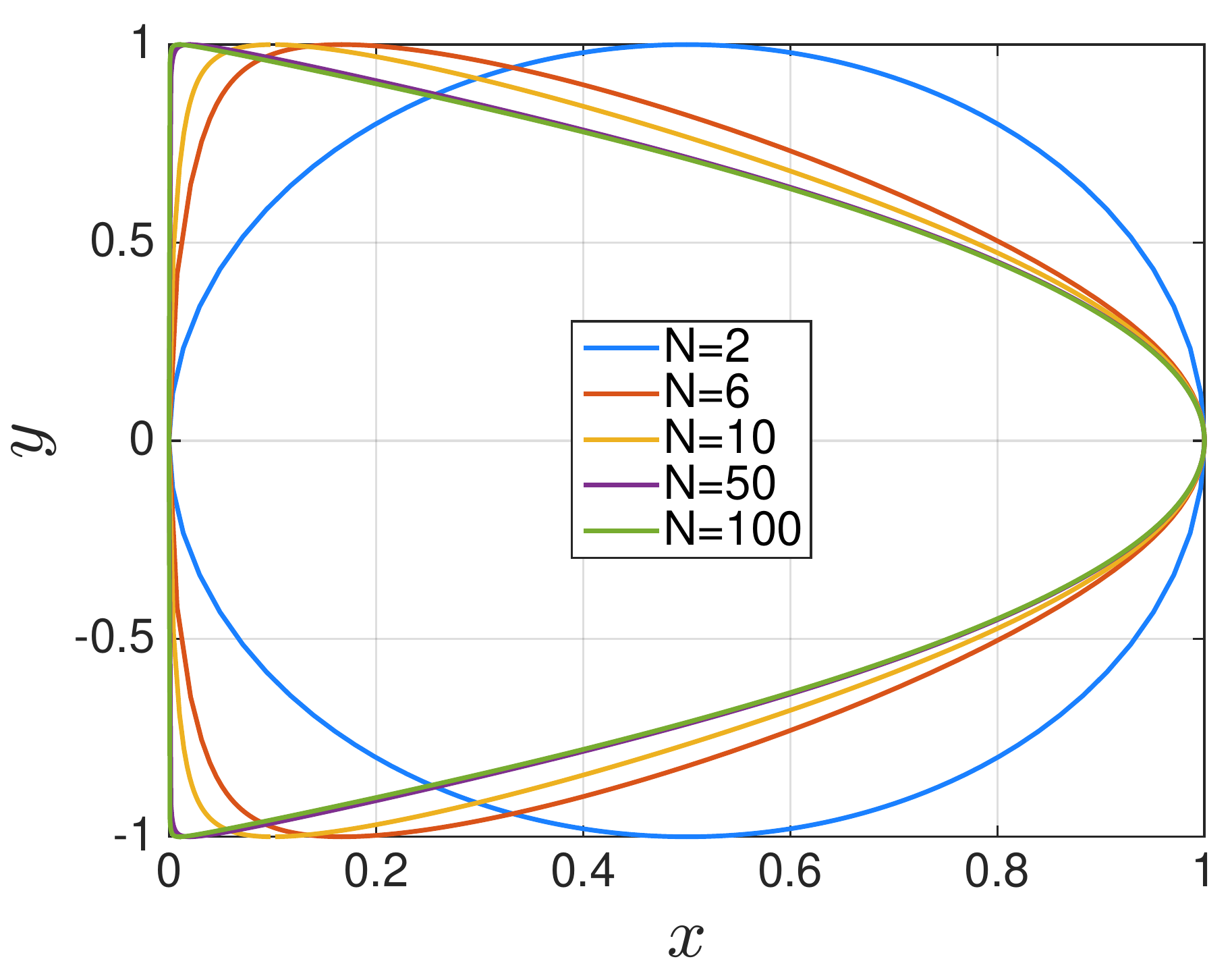}}\\
  \subfloat[]{
    \includegraphics[width=70mm,height=60mm]{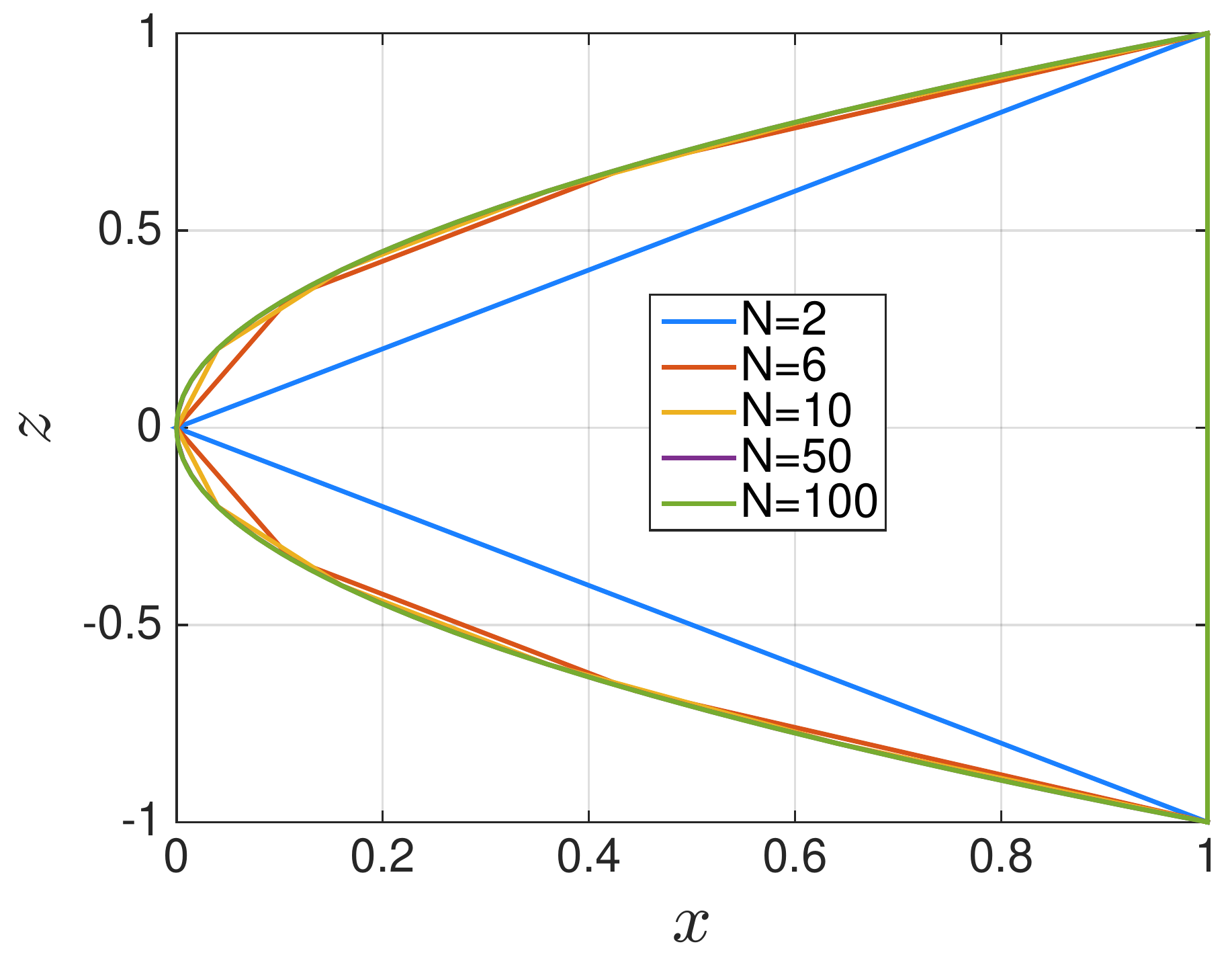}}\\
  \caption{2D projection of the two-mode Ising convex set $\Theta^{N}_2(H^{N}_{\mathrm{Ising}})$.
  $x=\tr\left(\rho^N_2(\frac{1}{N}I+\frac{N-1}{N}X\otimes X)\right)$,
  $y=\tr\left(\rho^N_2Z\otimes I\right)$,
  $z=\tr\left(\rho^N_2X\otimes I\right)$.
  (a) Projection on to the $xy$ plane. (b) Projection on to the $xz$ plane.}
  \label{Ising_projection}
\end{figure}

The different origins of the two ruled surfaces can also be
viewed from the two-dimensional projections of $\Theta^{N}_2(H^{N}_{\mathrm{Ising}})$,
as shown in Fig.~\ref{Ising_projection}.
In Fig.~\ref{Ising_projection}(a),
the projection is onto to the $xy$ plane, which corresponds to the Hamiltonian of
$B_x=0$ in $H^N_{\mathrm{Ising}}$. For $B_z=0$, the Hamiltonian becomes
\begin{equation}
H_{\text{Ising}}^N=\frac{J}{N}(J_x^N)^2,
\end{equation}
where is a constant that is independent of system size $N$.
The spectra of $H_{\text{Ising}}^N$ is then given by
\begin{equation}
E^N_k = \frac{J}{N}k^2, \ k=0,\pm 2, \pm 4, \ldots, \pm N.
\end{equation}
When $J>0$ (corresponding to the blue ruled surface), the ground state corresponding to $k=0$ is unique, and the spectra
is gapless for $N\rightarrow\infty$. When $J<0$ (corresponding to the green ruled surface),
the ground state corresponding to $k=\pm N$ is doubly degenerate, and
the spectra has a constant gap.

For any $N$ with $J>0$, the ground state of $\frac{J}{N}(J_x^N)^2$
is unique which is given by the eigenstate of $J_x^N$ with eigenvalue $0$ (denoted by
$\ket{J_x=0}$). This can be seen from the fact the line $x=0$ only intersects the
$2D$ projection of $\Theta^{N}_2(H^{N}_{\mathrm{Ising}})$
at the point $(0,0)$.

However, when $N\rightarrow\infty$, the eigenstates
of $J_z^N$ with the eigenvalue $\pm N$ (denoted by $\ket{J_z=\pm N}$) will have the same
energy per particle as that of $\ket{J_x=0}$. As a result, in the $N\rightarrow\infty$, the line
$x=0$ will intersect the convex set $\Theta^{\infty}_2(H^{\infty}_{\mathrm{Ising}})$
at an line interval with end points $(0,1)$ (corresponding to $\ket{J_z=N}$) and
$(0,-1)$ (corresponding to $\ket{J_z=-N}$). The behaviour of the curves approaching
the line interval for $x=0$ when $N$ increases clearly indicates a gapless system (hence
the origin of the blue ruled surface), together with
a first order phase transition at $B_z=0$ in the $N\rightarrow\infty$ limit.

A similar phenomenon can be observed for all the other line segments
on the blue ruled surface. For $J>0$,$B_x\neq 0$ and $B_z=0$, the Hamiltonian becomes
\begin{eqnarray}
H_{\text{Ising}}^N&=&\frac{J}{N}(J_x^N)^2+B_xJ_x^N\nonumber\\
&=&\frac{J}{N}\left(J_x^N+\frac{NB_x}{2J}\right)^2 - \frac{NB_x^2}{4J}.
\end{eqnarray}
The spectra of $H_{\text{Ising}}^N$ is then given by
\begin{equation}
E^N_k=\frac{J}{N}\left(k+\frac{NB_x}{2J}\right)^2 - \frac{NB_x^2}{4J},\ k=0,\pm 2, \pm 4, \ldots, \pm N
\end{equation}
which is also gapless for $N\rightarrow\infty$.

In Fig.~\ref{Ising_projection}(b),
the projection is onto to the $xz$ plane, which corresponds to the Hamiltonian of
$B_z=0$ in $H^N_{\mathrm{Ising}}$. For $J<0,B_x=0$, the ground state is two fold degenerate
that are given by $\ket{J_x=\pm N}$, which is even exact for finite $N$.
The system has a $\mathbb{Z}_2$ symmetry, and
the symmetry-breaking ground states are $\ket{J_x=\pm N}$. This
indicates that the green ruled surface is due to symmetry-breaking.

Before we end this section, we would like to remark that,
the stability of the points in the convex set is
not the same for all points. The points on the ruled
surfaces with gapless systems are more fragile than others, in the sense they only occur in
a relatively narrow parameter region and will leave that area under a
small parameter change. Points on the symmetry breaking ruled surface
are more stable. Also, there is some
`even-odd effect' for this model, i.e., for even particle number
and odd particle number, the result may have some difference. But this
difference is not important here, and in both cases the limit
will be the same $\Theta^{\infty}_2(H^{\infty}_{\mathrm{Ising}})$.
For simplicity we only show the $N$ even case.

\section{Gapless systems and ruled surfaces}

As discussed in Sec.~\ref{sec:Ising}, ruled surfaces on the
boundary of $\Theta_2^{\infty}(H^{\infty})$ may be a result of either
symmetry-breaking or gapless systems. If we know the gap/symmetry
properties of the system, then we can tell the physical origin of each
ruled surface. However, suppose we have no such knowledge of the
system and hope to learn something solely from the geometry, then
there is no way to tell such a difference (e.g. the blue and green
ruled surfaces in Fig.~\ref{Ising_infty} have exactly the same
geometric shape).

In order to tell the difference, we will then need to use finite size
scaling of $\Theta_2^{N}(H^{N})$. By computing boundary lines
corresponding to ground states of $H^{N}$, shown in Fig.~\ref{Ising_projection},
symmetry-breaking systems show very different behaviors comparing
to gapless systems. We will hence propose to use finite size scaling
to study the origin of ruled surfaces in $\Theta_2^{\infty}(H^{\infty})$,
and use the following system as an example for applying our idea of finite size scaling.

Consider the two-mode $\mathrm{XY}$ model, where $S_x=X,S_y=Y,S_z=Z$
are Pauli operators for qubit. The Hamiltonian reads
\begin{equation}
  H^N_{\mathrm{XY}}=\frac{1}{N} \left(J_1 (J^N_x)^2+J_2 (J^N_y)^2\right)+B_z J^N_z.
\end{equation}
The corresponding $\Theta^N_2(H^N_{\mathrm{XY}})$ is generated by
\begin{eqnarray}
  x &=& \frac{1}{N^2}\tr\left(\rho^N_2 (J^N_x)^2\right) \nonumber\\
    &=& \tr\left(\rho^N_2(\frac{1}{N}I+\frac{N-1}{N} X\otimes X)\right),\nonumber\\
  y &=& \frac{1}{N^2}\tr\left(\rho^N_2 (J^N_y)^2\right) \nonumber\\
    &=& \tr\left(\rho^N_2(\frac{1}{N}I+\frac{N-1}{N} Y\otimes Y)\right),\nonumber\\
  z &=& \frac{1}{N}\tr\left(\rho^N_2 J^N_z\right) = \tr\left(\rho^N_2Z\otimes I\right).\nonumber
\end{eqnarray}
for $\rho^N_2\in\Theta^N_2(H^N_{\mathrm{XY}})$. In the $N\rightarrow\infty$
limit, this is equivalent to
\begin{eqnarray}
  \Theta^{\infty}_2(H_{\mathrm{XY}}^{\infty})=&&\{\tr(\rho^{\infty}_2(X\otimes X)),\tr(\rho^{\infty}_2(Y\otimes Y)),\nonumber\\
                            &&\tr(\rho^{\infty}_2Z\otimes I))|\rho_2^{\infty}\in\Theta^{\infty}_2\}.
\end{eqnarray}
Let
\begin{eqnarray}
  \bar{x}&=&\tr(\rho^{\infty}_2(X\otimes X)),\nonumber\\
  \ \bar{y}&=&\tr(\rho^{\infty}_2(Y\otimes Y)),\\
  \ \bar{z}&=&\tr(\rho^{\infty}_2(Z\otimes I)),\nonumber
\end{eqnarray}
the extreme points of $\Theta^{\infty}_2(H^{\infty}_{\mathrm{XY}})$ satisfy
\begin{equation}\label{XYConvex}
  \bar{x}+\bar{y}+\bar{z}^2=1,\ \bar{x}\geq 0,\ \bar{y}\geq 0.
\end{equation}
This is also the boundary surface of $\Theta^{\infty}_2(H^{\infty}_{\mathrm{XY}})$.
The corresponding supporting hyperplanes are
\begin{equation}
  \bar{x}+\bar{x}_0+\bar{y}+\bar{y}_0+2\bar{z}_0\bar{z}=2,\ \bar{x}\geq 0,\ \bar{y}\geq 0.
\end{equation}

\begin{figure}[hbpt]
  \centering
  \includegraphics[width=80mm,height=70mm]{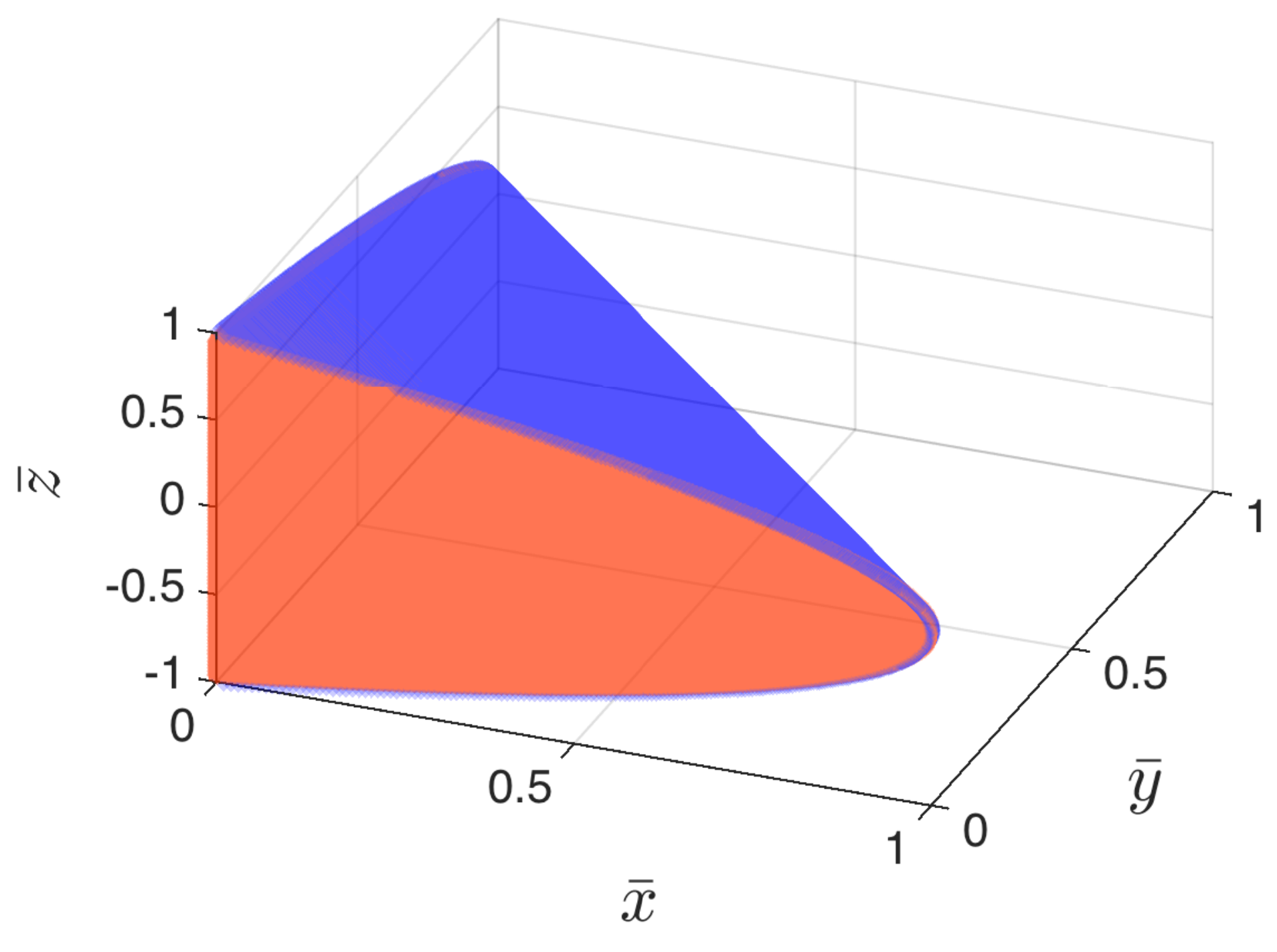}
  \caption{Convex set for two-mode $\mathrm{XY}$ model in the large $N$
    limit. It is determined by equation \eqref{XYConvex}, with $\bar{x}=\tr(\rho^{\infty}_2(X\otimes X))$,
   $\bar{y}=\tr(\rho^{\infty}_2(Y\otimes Y))$,
   $\bar{z}=\tr(\rho^{\infty}_2(Z\otimes I))$. The convex set is symmetric with
   respect to interchange of $\bar{x}$ and $\bar{y}$ axis. }
  \label{XY_infty}
\end{figure}

We show the convex set
$\Theta_2^{\infty}(H^{\infty}_{\mathrm{XY}})$ in Fig.~\ref{XY_infty}.
 There is a blued ruled surface on the boundary, together
 with two plane areas given by the intersection
 of $\Theta_2^{\infty}(H^{\infty}_{\mathrm{XY}})$ with the planes
 $\bar{x}=0$ and $\bar{y}=0$ respectively.
 For any point $(x_0,y_0,z_0)$ living on the surface
\begin{equation}\nonumber
  \bar{x}+\bar{y}+\bar{z}^2=1,\ \bar{x}\geq 0,\ \bar{y}\geq 0,
\end{equation}
part of the line $(x,1-x-z_0^2,z_0)$ also lives on this surface.

\begin{figure*}[hbpt]
  \centering
  \subfloat[]{
    \includegraphics[width=80mm,height=65mm]{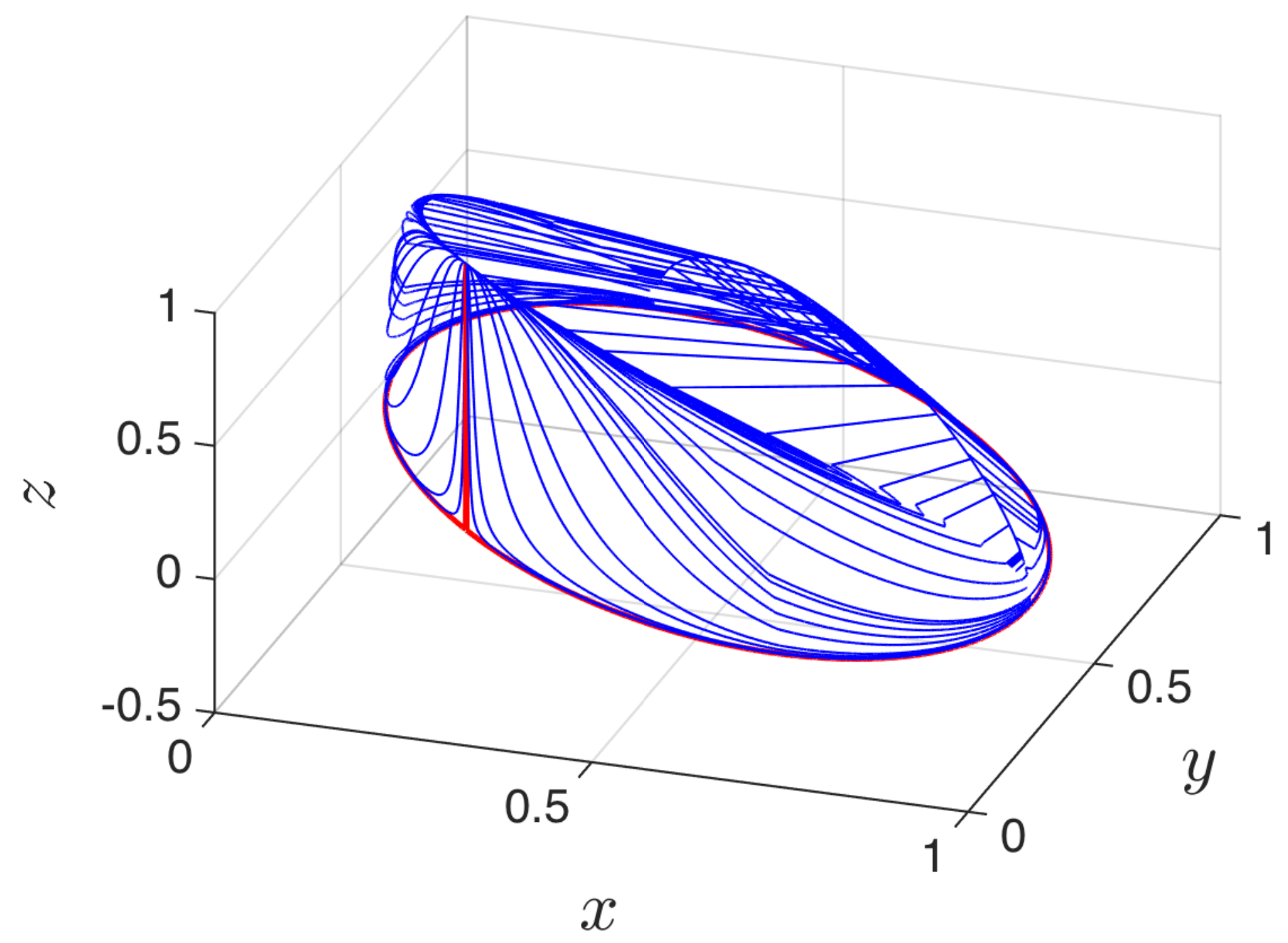}}
  \subfloat[]{
    \includegraphics[width=80mm,height=65mm]{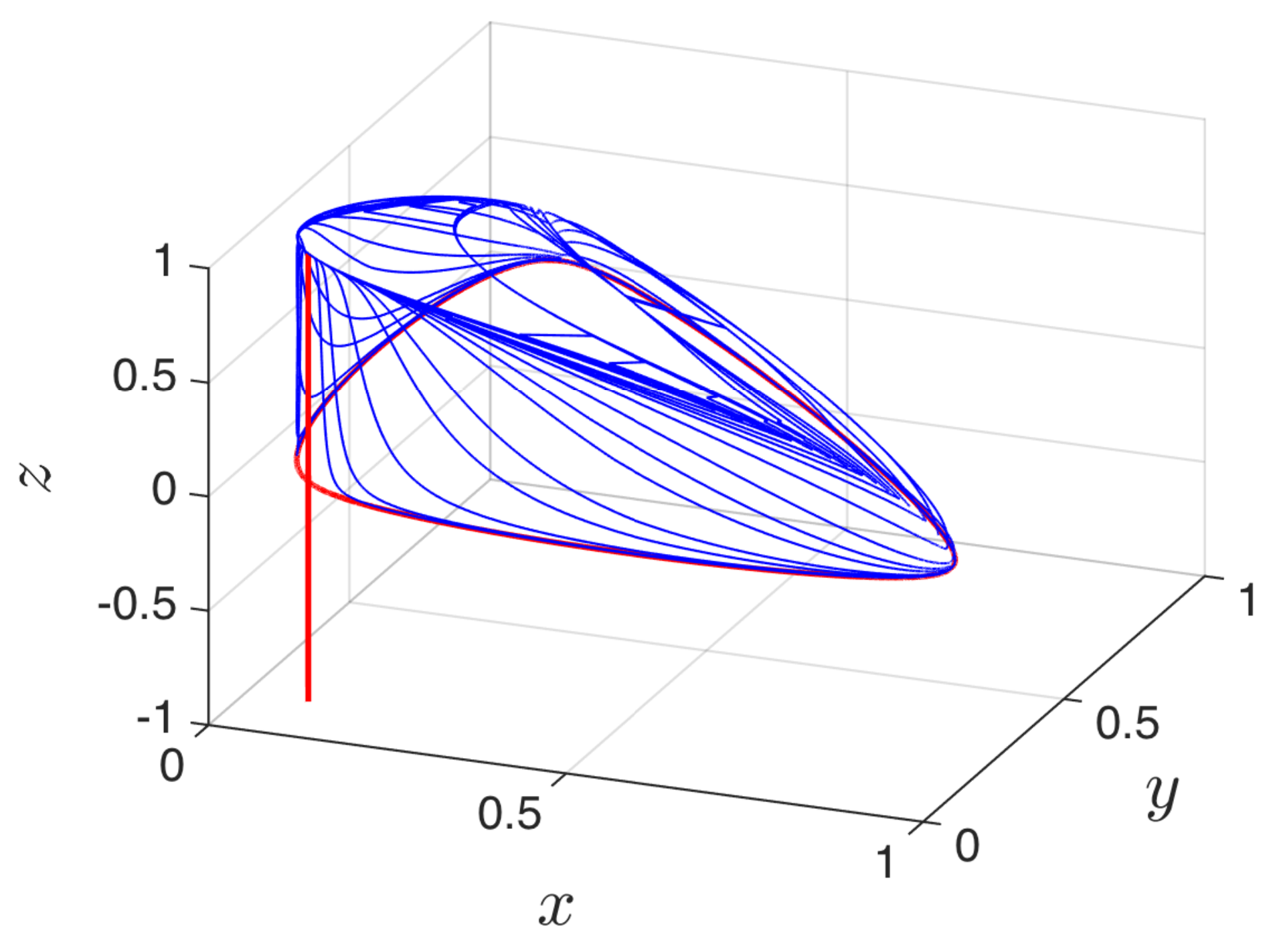}}\\
   \subfloat[]{
    \includegraphics[width=80mm,height=65mm]{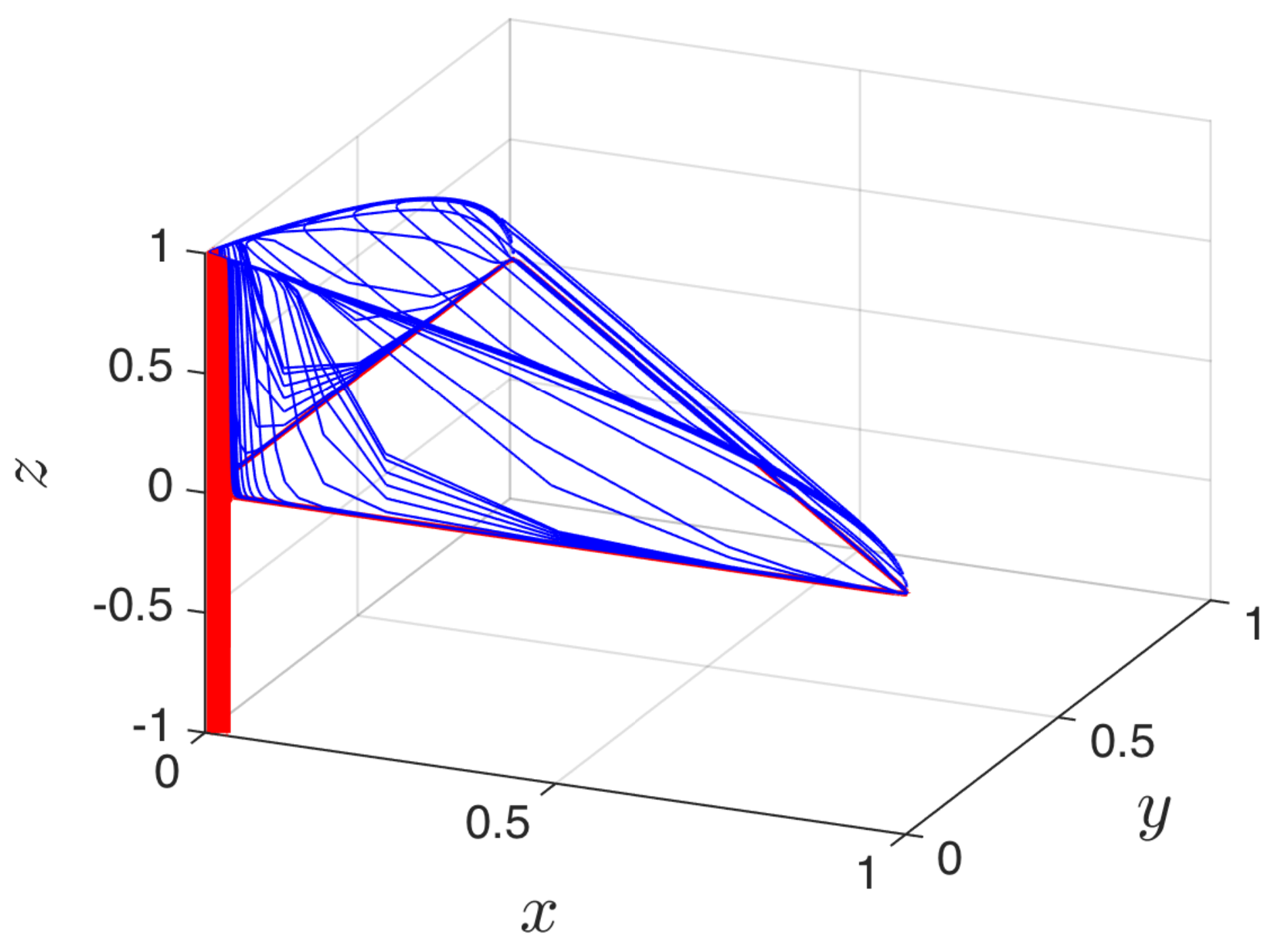}}
  \subfloat[]{
    \includegraphics[width=80mm,height=65mm]{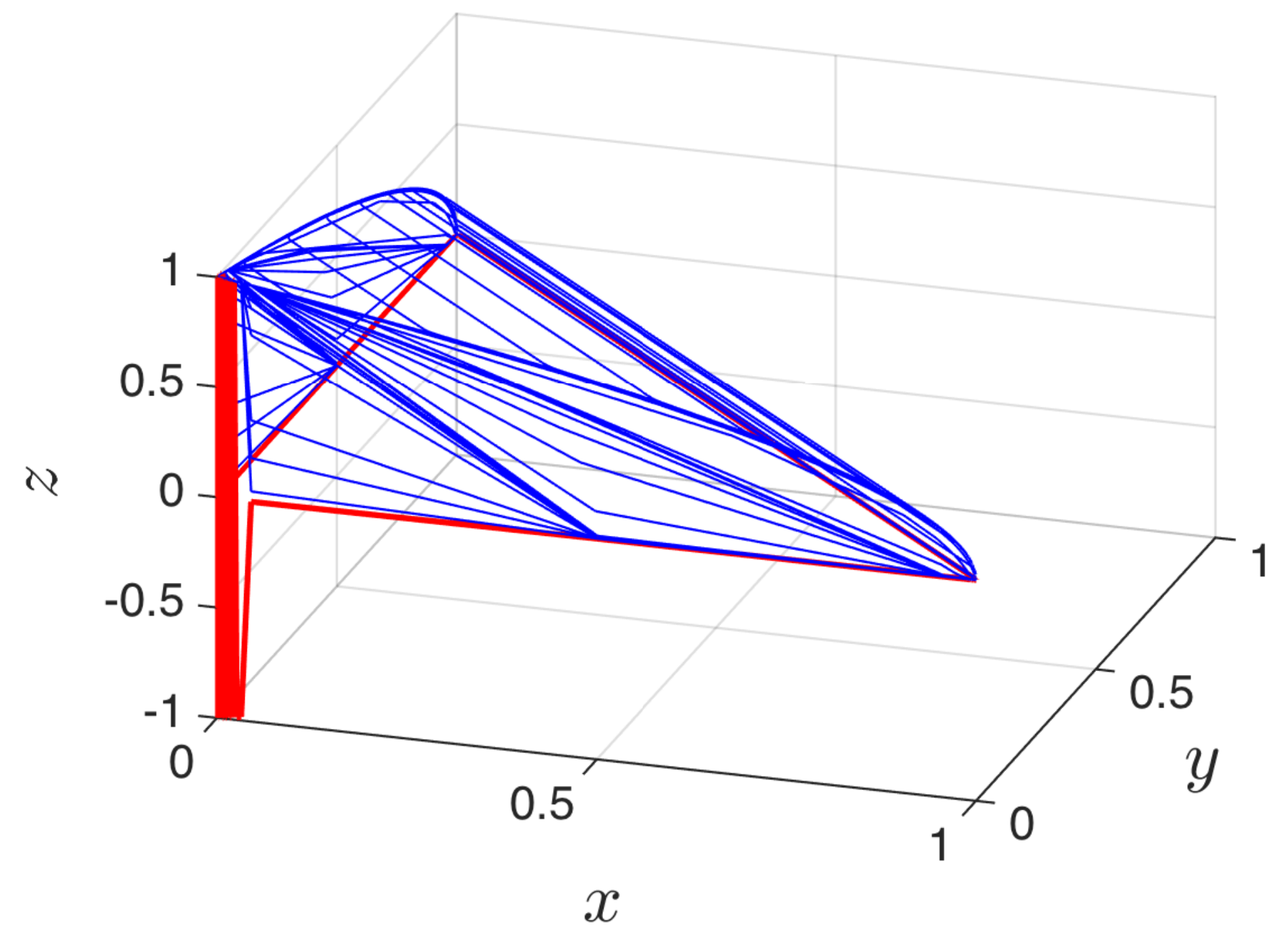}}
  \caption{Convex set $\Theta_2^{\infty}(H^{\infty}_{\mathrm{XY}})$ for two-mode $\mathrm{XY}$ model with finite system size $N$, where
    $x=\tr\left(\rho^N_2(\frac{1}{N}I+\frac{N-1}{N} X\otimes X)\right)$,
  $y=\tr\left(\rho^N_2(\frac{1}{N}I+\frac{N-1}{N} Y\otimes Y)\right)$,
  $z=\tr\left(\rho^N_2Z\otimes I\right)$.
    System size is $N=4,10,2\times10^2,10^3$, for (a),(b),(c),(d). For simplicity, we only show
    $B_z\le 0$ part.}
  \label{XY}
\end{figure*}

\begin{figure}[hbpt]
  \centering
  \includegraphics[width=70mm,height=60mm]{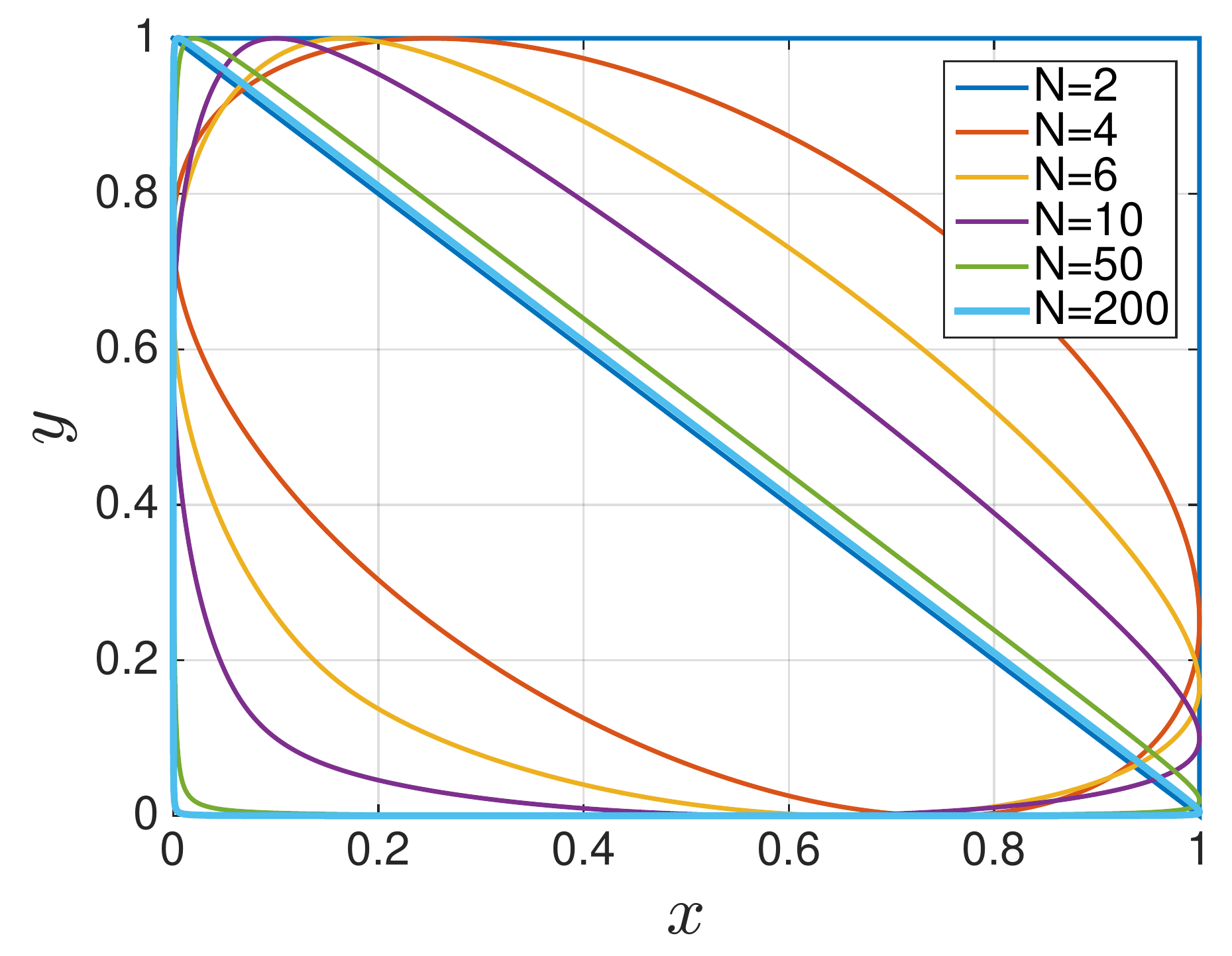}
  \caption{2D projection of $\Theta_2^{N}(H^{N}_{\mathrm{XY}})$ onto the $xy$ plane, where
  $x=\tr\left(\rho^N_2(\frac{1}{N}I+\frac{N-1}{N} X\otimes X)\right)$, $y=\tr\left(\rho^N_2(\frac{1}{N}I+\frac{N-1}{N} Y\otimes Y)\right)$.}
  \label{XY_projection}
\end{figure}

While the two planes corresponds to gapless systems,
the question is what is the origin of the blued ruled surfaces,
i.e., whether it results from symmetry-breaking or gapless systems.
To learn more information, we will then need the finite size scaling behaviors
of $\Theta_2^{N}(H^{N}_{\mathrm{XY}})$,
which we show in Fig.~\ref{XY}.

We also show the
$2D$ projection of  $\Theta_2^{N}(H^{N}_{\mathrm{XY}})$ onto the
$xy$ plane in Fig.~\ref{XY_projection}. This corresponds to $B_z=0$,
and the Hamiltonian becomes
\begin{equation}
  H^N_{\mathrm{XY}}=\frac{1}{N} \left(J_1 (J^N_x)^2+J_2 (J^N_y)^2\right).
\end{equation}
For $J_1=J_2$, we have
\begin{eqnarray}
  H^N_{\mathrm{XY}}&=&\frac{1}{N} \left(J_1 (J^N_x)^2+J_2 (J^N_y)^2\right)\nonumber\\
  &=&\frac{J_1}{N}((J^N)^2-(J^N_z)^2),
\end{eqnarray}
where the operator $(J^N)^2=(J^N_x)^2+(J^N_y)^2+(J^N_z)^2$.
The spectra of $H_{\mathrm{XY}}^N$ is then given by
\begin{equation}
E^N_k=\frac{J_1}{N}\left(N(N+2)-k^2\right),\ k=0,\pm 2,\pm 4,\ldots, \pm N,
\end{equation}
which is gapless for $N\rightarrow\infty$ when $J_1<0$ (corresponding
to the blue ruled surface).

The projection of
$\Theta_2^{\infty}(H^{\infty}_{\mathrm{XY}})$ is in fact
a triangle with vertices $(0,0),(1,0),(0,1)$. The finite size
scaling of $\Theta_2^{N}(H^{N}_{\mathrm{XY}})$ clearly approaches
each boundary line of this triangle in the $N\rightarrow\infty$
limit, indicating gapless systems. That is, the blue
ruled surface results from gapless systems. There
is in fact no ruled surface resulting from symmetry-breaking
in the geometry of $\Theta_2^{N}(H^{N}_{\mathrm{XY}})$.

\section{Discussion}

In this work, we have examined the geometry of reduced density
matrices for bosonic systems, which are convex sets in $\mathbb{R}^3$.
Our focus is on the physical origin of the ruled surfaces on the
boundary of these convex sets. We show that apart from signatures of
symmetry-breaking, ruled surfaces can also be a consequence of gapless
systems. Concrete examples are examined for bosonic system in the
$N\rightarrow\infty$ limit, and ruled surfaces due to gapless systems
are shown. Thanks to the quantum de Finetti's theorem, the geometry of
the reduced density matrices of the discussed bosonic models can be
found analytically.

In more general cases where there is no longer bosonic exchange
symmetry, we would expect that the relationship between gapless
systems and the emergence of ruled surface on the boundary of
three-dimensional projections of reduced density matrices will remain
valid, since the bosonic exchange symmetry is not essential for having
those ruled surfaces. For general systems without bosonic exchange
symmetry, however, quantum de Finetti's theorem is no longer valid,
and the geometry of $2$-RDMs is hard to get in general~\cite{Liu06,LCV07,WMN10}.
Nevertheless, it is interesting to study other concrete systems whose
geometry of $2$-RDMs will also have ruled surface on the boundary that
is related to gapless systems. We leave this for future work.

\section*{Acknowledgement} We thank Yi Shen and Li You for helpful
discussions. B.Z. is supported by NSERC and CIFAR.
X.Q. is supported by program for the Outstanding Innovative Teams of Higher Learning Institutions of Shanxi.
This research was supported in part by Perimeter Institute for Theoretical Physics.
Research at Perimeter Institute is supported by the Government of Canada through Industry Canada and by the Province
of Ontario through the Ministry of Economic Development \& Innovation.

\bibliography{Boson}

\begin{thebibliography}{21}%
\makeatletter
\providecommand \@ifxundefined [1]{%
 \@ifx{#1\undefined}
}%
\providecommand \@ifnum [1]{%
 \ifnum #1\expandafter \@firstoftwo
 \else \expandafter \@secondoftwo
 \fi
}%
\providecommand \@ifx [1]{%
 \ifx #1\expandafter \@firstoftwo
 \else \expandafter \@secondoftwo
 \fi
}%
\providecommand \natexlab [1]{#1}%
\providecommand \enquote  [1]{``#1''}%
\providecommand \bibnamefont  [1]{#1}%
\providecommand \bibfnamefont [1]{#1}%
\providecommand \citenamefont [1]{#1}%
\providecommand \href@noop [0]{\@secondoftwo}%
\providecommand \href [0]{\begingroup \@sanitize@url \@href}%
\providecommand \@href[1]{\@@startlink{#1}\@@href}%
\providecommand \@@href[1]{\endgroup#1\@@endlink}%
\providecommand \@sanitize@url [0]{\catcode `\\12\catcode `\$12\catcode
  `\&12\catcode `\#12\catcode `\^12\catcode `\_12\catcode `\%12\relax}%
\providecommand \@@startlink[1]{}%
\providecommand \@@endlink[0]{}%
\providecommand \url  [0]{\begingroup\@sanitize@url \@url }%
\providecommand \@url [1]{\endgroup\@href {#1}{\urlprefix }}%
\providecommand \urlprefix  [0]{URL }%
\providecommand \Eprint [0]{\href }%
\providecommand \doibase [0]{http://dx.doi.org/}%
\providecommand \selectlanguage [0]{\@gobble}%
\providecommand \bibinfo  [0]{\@secondoftwo}%
\providecommand \bibfield  [0]{\@secondoftwo}%
\providecommand \translation [1]{[#1]}%
\providecommand \BibitemOpen [0]{}%
\providecommand \bibitemStop [0]{}%
\providecommand \bibitemNoStop [0]{.\EOS\space}%
\providecommand \EOS [0]{\spacefactor3000\relax}%
\providecommand \BibitemShut  [1]{\csname bibitem#1\endcsname}%
\let\auto@bib@innerbib\@empty
\bibitem [{\citenamefont {Coleman}(1963)}]{Col63}%
  \BibitemOpen
  \bibfield  {author} {\bibinfo {author} {\bibfnamefont {A.~J.}\ \bibnamefont
  {Coleman}},\ }\href {\doibase 10.1103/RevModPhys.35.668} {\bibfield
  {journal} {\bibinfo  {journal} {Rev. Mod. Phys.}\ }\textbf {\bibinfo {volume}
  {35}},\ \bibinfo {pages} {668} (\bibinfo {year} {1963})}\BibitemShut
  {NoStop}%
\bibitem [{\citenamefont {Erdahl}(1972)}]{Erd72}%
  \BibitemOpen
  \bibfield  {author} {\bibinfo {author} {\bibfnamefont {R.~M.}\ \bibnamefont
  {Erdahl}},\ }\href {\doibase http://dx.doi.org/10.1063/1.1665885} {\bibfield
  {journal} {\bibinfo  {journal} {Journal of Mathematical Physics}\ }\textbf
  {\bibinfo {volume} {13}},\ \bibinfo {pages} {1608} (\bibinfo {year}
  {1972})}\BibitemShut {NoStop}%
\bibitem [{\citenamefont {Klyachko}(2006)}]{klyachko2006quantum}%
  \BibitemOpen
  \bibfield  {author} {\bibinfo {author} {\bibfnamefont {A.~A.}\ \bibnamefont
  {Klyachko}},\ }in\ \href@noop {} {\emph {\bibinfo {booktitle} {Journal of
  Physics: Conference Series}}},\ Vol.~\bibinfo {volume} {36}\ (\bibinfo
  {organization} {IOP Publishing},\ \bibinfo {year} {2006})\ p.~\bibinfo
  {pages} {72}\BibitemShut {NoStop}%
\bibitem [{\citenamefont {Erdahl}\ and\ \citenamefont {Jin}(2000)}]{EJ00}%
  \BibitemOpen
  \bibfield  {author} {\bibinfo {author} {\bibfnamefont {R.}~\bibnamefont
  {Erdahl}}\ and\ \bibinfo {author} {\bibfnamefont {B.}~\bibnamefont {Jin}},\
  }in\ \href {\doibase 10.1007/978-1-4615-4211-7_4} {\emph {\bibinfo
  {booktitle} {Many-Electron Densities and Reduced Density Matrices}}},\
  \bibinfo {series and number} {Mathematical and Computational Chemistry},\
  \bibinfo {editor} {edited by\ \bibinfo {editor} {\bibfnamefont
  {J.}~\bibnamefont {Cioslowski}}}\ (\bibinfo  {publisher} {Springer US},\
  \bibinfo {year} {2000})\ pp.\ \bibinfo {pages} {57--84}\BibitemShut {NoStop}%
\bibitem [{\citenamefont {Schwerdtfeger}\ and\ \citenamefont
  {Mazziotti}(2009)}]{SM09}%
  \BibitemOpen
  \bibfield  {author} {\bibinfo {author} {\bibfnamefont {C.~A.}\ \bibnamefont
  {Schwerdtfeger}}\ and\ \bibinfo {author} {\bibfnamefont {D.~A.}\ \bibnamefont
  {Mazziotti}},\ }\href {\doibase http://dx.doi.org/10.1063/1.3143403}
  {\bibfield  {journal} {\bibinfo  {journal} {The Journal of Chemical Physics}\
  }\textbf {\bibinfo {volume} {130}},\ \bibinfo {eid} {224102} (\bibinfo {year}
  {2009})}\BibitemShut {NoStop}%
\bibitem [{\citenamefont {Liu}(2006)}]{Liu06}%
  \BibitemOpen
  \bibfield  {author} {\bibinfo {author} {\bibfnamefont {Y.-K.}\ \bibnamefont
  {Liu}},\ }in\ \href {\doibase 10.1007/11830924_40} {\emph {\bibinfo
  {booktitle} {Approximation, Randomization, and Combinatorial Optimization.
  Algorithms and Techniques}}},\ \bibinfo {series} {Lecture Notes in Computer
  Science}, Vol.\ \bibinfo {volume} {4110},\ \bibinfo {editor} {edited by\
  \bibinfo {editor} {\bibfnamefont {J.}~\bibnamefont {Diaz}}, \bibinfo {editor}
  {\bibfnamefont {K.}~\bibnamefont {Jansen}}, \bibinfo {editor} {\bibfnamefont
  {J.~D.}\ \bibnamefont {Rolim}}, \ and\ \bibinfo {editor} {\bibfnamefont
  {U.}~\bibnamefont {Zwick}}}\ (\bibinfo  {publisher} {Springer Berlin
  Heidelberg},\ \bibinfo {year} {2006})\ pp.\ \bibinfo {pages}
  {438--449}\BibitemShut {NoStop}%
\bibitem [{\citenamefont {Liu}\ \emph {et~al.}(2007)\citenamefont {Liu},
  \citenamefont {Christandl},\ and\ \citenamefont {Verstraete}}]{LCV07}%
  \BibitemOpen
  \bibfield  {author} {\bibinfo {author} {\bibfnamefont {Y.-K.}\ \bibnamefont
  {Liu}}, \bibinfo {author} {\bibfnamefont {M.}~\bibnamefont {Christandl}}, \
  and\ \bibinfo {author} {\bibfnamefont {F.}~\bibnamefont {Verstraete}},\
  }\href {\doibase 10.1103/PhysRevLett.98.110503} {\bibfield  {journal}
  {\bibinfo  {journal} {Phys. Rev. Lett.}\ }\textbf {\bibinfo {volume} {98}},\
  \bibinfo {pages} {110503} (\bibinfo {year} {2007})}\BibitemShut {NoStop}%
\bibitem [{\citenamefont {Wei}\ \emph {et~al.}(2010)\citenamefont {Wei},
  \citenamefont {Mosca},\ and\ \citenamefont {Nayak}}]{WMN10}%
  \BibitemOpen
  \bibfield  {author} {\bibinfo {author} {\bibfnamefont {T.-C.}\ \bibnamefont
  {Wei}}, \bibinfo {author} {\bibfnamefont {M.}~\bibnamefont {Mosca}}, \ and\
  \bibinfo {author} {\bibfnamefont {A.}~\bibnamefont {Nayak}},\ }\href
  {\doibase 10.1103/PhysRevLett.104.040501} {\bibfield  {journal} {\bibinfo
  {journal} {Phys. Rev. Lett.}\ }\textbf {\bibinfo {volume} {104}},\ \bibinfo
  {pages} {040501} (\bibinfo {year} {2010})}\BibitemShut {NoStop}%
\bibitem [{\citenamefont {Verstraete}\ and\ \citenamefont
  {Cirac}(2006)}]{VC06}%
  \BibitemOpen
  \bibfield  {author} {\bibinfo {author} {\bibfnamefont {F.}~\bibnamefont
  {Verstraete}}\ and\ \bibinfo {author} {\bibfnamefont {J.~I.}\ \bibnamefont
  {Cirac}},\ }\href {\doibase 10.1103/PhysRevB.73.094423} {\bibfield  {journal}
  {\bibinfo  {journal} {Phys. Rev. B}\ }\textbf {\bibinfo {volume} {73}},\
  \bibinfo {pages} {094423} (\bibinfo {year} {2006})}\BibitemShut {NoStop}%
\bibitem [{\citenamefont {Gidofalvi}\ and\ \citenamefont
  {Mazziotti}(2006)}]{GM06}%
  \BibitemOpen
  \bibfield  {author} {\bibinfo {author} {\bibfnamefont {G.}~\bibnamefont
  {Gidofalvi}}\ and\ \bibinfo {author} {\bibfnamefont {D.~A.}\ \bibnamefont
  {Mazziotti}},\ }\href {\doibase 10.1103/PhysRevA.74.012501} {\bibfield
  {journal} {\bibinfo  {journal} {Phys. Rev. A}\ }\textbf {\bibinfo {volume}
  {74}},\ \bibinfo {pages} {012501} (\bibinfo {year} {2006})}\BibitemShut
  {NoStop}%
\bibitem [{\citenamefont {Chen}\ \emph {et~al.}(2015)\citenamefont {Chen},
  \citenamefont {Ji}, \citenamefont {Li}, \citenamefont {Poon}, \citenamefont
  {Shen}, \citenamefont {Yu}, \citenamefont {Zeng},\ and\ \citenamefont
  {Zhou}}]{chen2015discontinuity}%
  \BibitemOpen
  \bibfield  {author} {\bibinfo {author} {\bibfnamefont {J.}~\bibnamefont
  {Chen}}, \bibinfo {author} {\bibfnamefont {Z.}~\bibnamefont {Ji}}, \bibinfo
  {author} {\bibfnamefont {C.-K.}\ \bibnamefont {Li}}, \bibinfo {author}
  {\bibfnamefont {Y.-T.}\ \bibnamefont {Poon}}, \bibinfo {author}
  {\bibfnamefont {Y.}~\bibnamefont {Shen}}, \bibinfo {author} {\bibfnamefont
  {N.}~\bibnamefont {Yu}}, \bibinfo {author} {\bibfnamefont {B.}~\bibnamefont
  {Zeng}}, \ and\ \bibinfo {author} {\bibfnamefont {D.}~\bibnamefont {Zhou}},\
  }\href@noop {} {\bibfield  {journal} {\bibinfo  {journal} {New Journal of
  Physics}\ }\textbf {\bibinfo {volume} {17}},\ \bibinfo {pages} {083019}
  (\bibinfo {year} {2015})}\BibitemShut {NoStop}%
\bibitem [{\citenamefont {Zauner}\ \emph {et~al.}(2014)\citenamefont {Zauner},
  \citenamefont {Vanderstraeten}, \citenamefont {Draxler}, \citenamefont
  {Lee},\ and\ \citenamefont {Verstraete}}]{zauner2014symmetry}%
  \BibitemOpen
  \bibfield  {author} {\bibinfo {author} {\bibfnamefont {V.}~\bibnamefont
  {Zauner}}, \bibinfo {author} {\bibfnamefont {L.}~\bibnamefont
  {Vanderstraeten}}, \bibinfo {author} {\bibfnamefont {D.}~\bibnamefont
  {Draxler}}, \bibinfo {author} {\bibfnamefont {Y.}~\bibnamefont {Lee}}, \ and\
  \bibinfo {author} {\bibfnamefont {F.}~\bibnamefont {Verstraete}},\
  }\href@noop {} {\bibfield  {journal} {\bibinfo  {journal} {arXiv preprint
  arXiv:1412.7642}\ } (\bibinfo {year} {2014})}\BibitemShut {NoStop}%
\bibitem [{\citenamefont {Chen}\ \emph {et~al.}(2016)\citenamefont {Chen},
  \citenamefont {Ji}, \citenamefont {Liu}, \citenamefont {Shen},\ and\
  \citenamefont {Zeng}}]{chen2016geometry}%
  \BibitemOpen
  \bibfield  {author} {\bibinfo {author} {\bibfnamefont {J.-Y.}\ \bibnamefont
  {Chen}}, \bibinfo {author} {\bibfnamefont {Z.}~\bibnamefont {Ji}}, \bibinfo
  {author} {\bibfnamefont {Z.-X.}\ \bibnamefont {Liu}}, \bibinfo {author}
  {\bibfnamefont {Y.}~\bibnamefont {Shen}}, \ and\ \bibinfo {author}
  {\bibfnamefont {B.}~\bibnamefont {Zeng}},\ }\href@noop {} {\bibfield
  {journal} {\bibinfo  {journal} {Physical Review A}\ }\textbf {\bibinfo
  {volume} {93}},\ \bibinfo {pages} {012309} (\bibinfo {year}
  {2016})}\BibitemShut {NoStop}%
\bibitem [{\citenamefont {Gibbs}(1873{\natexlab{a}})}]{Gib73a}%
  \BibitemOpen
  \bibfield  {author} {\bibinfo {author} {\bibfnamefont {J.~W.}\ \bibnamefont
  {Gibbs}},\ }\href@noop {} {\bibfield  {journal} {\bibinfo  {journal}
  {Transcations of the Connecticut Academy}\ }\textbf {\bibinfo {volume} {2}},\
  \bibinfo {pages} {309} (\bibinfo {year} {1873}{\natexlab{a}})}\BibitemShut
  {NoStop}%
\bibitem [{\citenamefont {Gibbs}(1873{\natexlab{b}})}]{Gib73b}%
  \BibitemOpen
  \bibfield  {author} {\bibinfo {author} {\bibfnamefont {J.~W.}\ \bibnamefont
  {Gibbs}},\ }\href@noop {} {\bibfield  {journal} {\bibinfo  {journal}
  {Transcations of the Connecticut Academy}\ }\textbf {\bibinfo {volume} {2}},\
  \bibinfo {pages} {382} (\bibinfo {year} {1873}{\natexlab{b}})}\BibitemShut
  {NoStop}%
\bibitem [{\citenamefont {Gibbs}(1875)}]{Gib75}%
  \BibitemOpen
  \bibfield  {author} {\bibinfo {author} {\bibfnamefont {J.~W.}\ \bibnamefont
  {Gibbs}},\ }\href@noop {} {\bibfield  {journal} {\bibinfo  {journal}
  {Transcations of the Connecticut Academy}\ }\textbf {\bibinfo {volume} {3}},\
  \bibinfo {pages} {108} (\bibinfo {year} {1875})}\BibitemShut {NoStop}%
\bibitem [{\citenamefont {Israel}(1979)}]{Isr79}%
  \BibitemOpen
  \bibfield  {author} {\bibinfo {author} {\bibfnamefont {R.~B.}\ \bibnamefont
  {Israel}},\ }\href@noop {} {\emph {\bibinfo {title} {Convexity in the Theory
  of Lattice Gases}}}\ (\bibinfo  {publisher} {Princeton University Press},\
  \bibinfo {year} {1979})\BibitemShut {NoStop}%
\bibitem [{\citenamefont {St{\o}rmer}(1969)}]{stormer1969symmetric}%
  \BibitemOpen
  \bibfield  {author} {\bibinfo {author} {\bibfnamefont {E.}~\bibnamefont
  {St{\o}rmer}},\ }\href@noop {} {\bibfield  {journal} {\bibinfo  {journal}
  {Journal of Functional Analysis}\ }\textbf {\bibinfo {volume} {3}},\ \bibinfo
  {pages} {48} (\bibinfo {year} {1969})}\BibitemShut {NoStop}%
\bibitem [{\citenamefont {Hudson}\ and\ \citenamefont
  {Moody}(1976)}]{hudson1976locally}%
  \BibitemOpen
  \bibfield  {author} {\bibinfo {author} {\bibfnamefont {R.~L.}\ \bibnamefont
  {Hudson}}\ and\ \bibinfo {author} {\bibfnamefont {G.~R.}\ \bibnamefont
  {Moody}},\ }\href@noop {} {\bibfield  {journal} {\bibinfo  {journal}
  {Probability Theory and Related Fields}\ }\textbf {\bibinfo {volume} {33}},\
  \bibinfo {pages} {343} (\bibinfo {year} {1976})}\BibitemShut {NoStop}%
\bibitem [{\citenamefont {Lewin}\ \emph {et~al.}(2014)\citenamefont {Lewin},
  \citenamefont {Nam},\ and\ \citenamefont {Rougerie}}]{lewin2014derivation}%
  \BibitemOpen
  \bibfield  {author} {\bibinfo {author} {\bibfnamefont {M.}~\bibnamefont
  {Lewin}}, \bibinfo {author} {\bibfnamefont {P.~T.}\ \bibnamefont {Nam}}, \
  and\ \bibinfo {author} {\bibfnamefont {N.}~\bibnamefont {Rougerie}},\
  }\href@noop {} {\bibfield  {journal} {\bibinfo  {journal} {Advances in
  Mathematics}\ }\textbf {\bibinfo {volume} {254}},\ \bibinfo {pages} {570}
  (\bibinfo {year} {2014})}\BibitemShut {NoStop}%
\bibitem [{\citenamefont {Leggett}(2001)}]{leggett2001bose}%
  \BibitemOpen
  \bibfield  {author} {\bibinfo {author} {\bibfnamefont {A.~J.}\ \bibnamefont
  {Leggett}},\ }\href@noop {} {\bibfield  {journal} {\bibinfo  {journal}
  {Reviews of Modern Physics}\ }\textbf {\bibinfo {volume} {73}},\ \bibinfo
  {pages} {307} (\bibinfo {year} {2001})}\BibitemShut {NoStop}%
\end{thebibliography}%

\end{document}